\documentclass{PoS}
\usepackage{amsfonts}%
\usepackage{mathrsfs}     
\title{Small, dark, and heavy: But is it a black hole?}

\ShortTitle{Small, dark, and heavy: But is it a black hole?}

\author{\speaker{Matt Visser}%
        \thanks{This research was supported by the Marsden Fund administered by the Royal Society of New Zealand}\\
       School of Mathematics, Statistics, and Operations Research,
Victoria University of Wellington, New Zealand\\
       E-mail: \email{matt.visser@msor.vuw.ac.nz}}
       
\author{ {Carlos Barcel\'o}\\
          Instituto de Astrof\'{\i}sica de Andaluc\'{\i}a, CSIC,
          Camino Bajo de Hu\'etor 50, 18008 Granada, Spain\\
            E-mail: \email{carlos@iaa.es}}     
          
\author{Stefano Liberati\\
            International School for Advanced Studies (SISSA), 
            Via Beirut 2-4, 34014 Trieste, Italy \\
            and INFN, Sezione di Trieste\\
             E-mail: \email{liberati@sissa.it}}      
            
\author{Sebastiano Sonego\\
             Universit\`a di Udine, Via delle Scienze 208, 33100 Udine, Italy\\
              E-mail: \email{sebastiano.sonego@uniud.it}}

\abstract{Astronomers have certainly observed things that are 
small, dark, and heavy. But are these objects really black holes 
in the sense of general relativity?  The consensus opinion is simply  
``yes'', and there is very little ``wriggle room''.  We discuss one of the specific alternatives.

\vskip 20pt

1 February 2009; 10 February 2009; \LaTeX-ed \today}

\FullConference{Black Holes in General Relativity and String Theory\\
		 24--30 August, 2008\\
		 Veli Losinj, Croatia}

\begin{document}
\newcommand{\scri}{\mathscr{I}}
\def\e{{\mathrm e}}%
\def\g{{\mbox{\sl g}}}%
\def\Box{\nabla^2}%
\def\d{{\mathrm d}}%
\def\R{{\rm I\!R}}%
\def\ie{{\em i.e.\/}}%
\def\eg{{\em e.g.\/}}%
\def\etc{{\em etc.\/}}%
\def\etal{{\em et al.\/}}%
\def\implies{{\Rightarrow}}

\section{Introduction}

\noindent
Do alternatives to standard classical black holes exist? Can one ``mimic'' a black hole with arbitrary accuracy? There is a rather limited set of (arguably) viable alternatives:
\begin{itemize}
\item 
Quark stars~\cite{quark}, Q-balls~\cite{q-ball}, strange stars~\cite{strange}.
\item
Boson-stars~\cite{boson}.
\item
Gravastars: Mazur--Mottola variant~\cite{Mazur-Mottola},  and  Laughlin \emph{et al.\/}~variant~\cite{Laughlin}.
\item
Fuzz-balls: Mathur \emph{et al.\/}~variant~\cite{Mathur}, and Amati variant~\cite{Amati}. (See~\cite{Skenderis} for a survey.)
\item
Dark stars/quasi-black holes~\cite{fate,fate2}. (For related ideas, see~\cite{Boulware, Vachaspati}).
\end{itemize}
While there are close inter-relationships between these various models, in this article we will specifically focus on our own proposal~\cite{fate,fate2}, and give an informal overview of the situation.  We shall re-assess and (hopefully) re-invigorate an old line of argument: What effect does quantum physics have on the collapse of a classical star?  Is semi-classical collapse~\cite{fate, Boulware, Vachaspati, Hajicek, Hajicek2} qualitatively different from classical collapse~\cite{Harrison-et-al}?

In general we can certainly write
\begin{equation}
G_{ab} = 8\pi  \langle\psi |\hat T_{ab} |\psi\rangle\;, 
\end{equation}
and separate the expectation value of the stress-energy-momentum  tensor into a contribution from some suitably chosen vacuum state, plus a contribution from the excitations above that vacuum state.  For instance, for an uncollapsed star
\begin{equation}
G_{ab} = 8\pi \left( T_{ab}^\mathrm{classical} + \langle 0|\hat T_{ab} |0\rangle\right)\;.
\label{semicl}
\end{equation}
The vacuum polarization effect $\langle 0 | \hat T_{ab} |0\rangle$ is utterly negligible in an ordinary uncollapsed star.  (This, after all,  is why we can get away with just solving the \emph{classical}\, Einstein equations most of the time.)  Does this remain true during collapse?  Even if the vacuum polarization does remain small, it might still have a significant effect on the location and/or existence of event horizons~\cite{Hajicek, Hajicek2}.

Now this point of view, while certainly historically respectable, does deviate significantly from the present ``consensus opinion'', at least in the general relativity community, so before one gets started there are a number of preliminary issues that should be dealt with.

\section{The Fulling--Sweeny--Wald (no-singularity) theorem}

\begin{figure}[!htbp]
\begin{center}
\includegraphics[width=8cm]{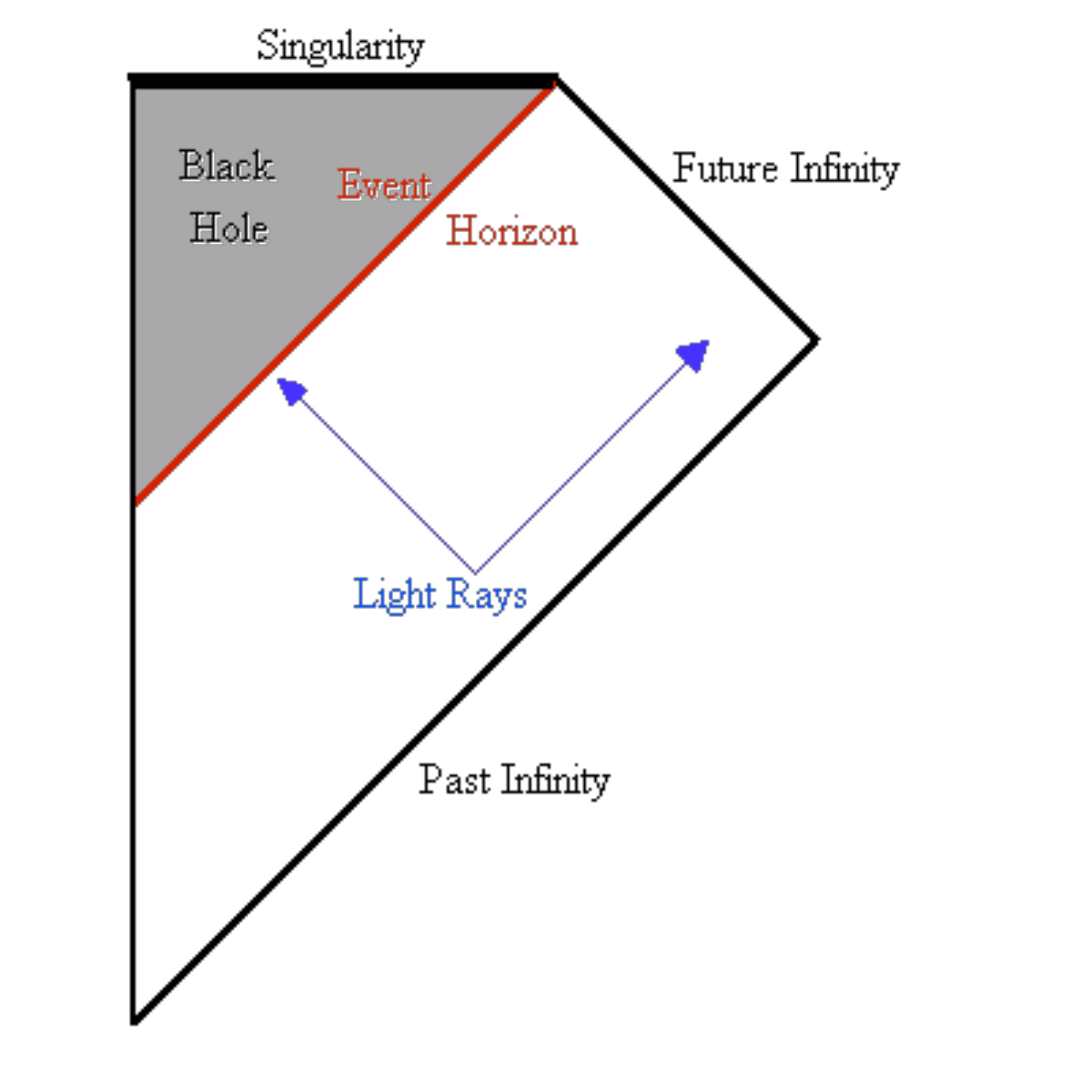}
\caption{Standard Carter--Penrose diagram for an astrophysical black hole formed via stellar collapse.}
\label{F:strict}
\end{center}
\end{figure}

There is a widespread feeling in the general relativity community that semiclassical quantum back-reaction effects are \emph{always\/} small, and \emph{never\/} enough to significantly alter the classical picture of collapse to a black hole.  (See figure~\ref{F:strict} for an appropriate Carter--Penrose diagram.) When pushed, members of this community ultimately point to the Fulling--Sweeny--Wald no-singularity theorem~\cite{FSW} as the basis for this assertion.

Phrasing the Fulling--Sweeny--Wald theorem rather loosely:  ``In quantum field theory on a curved spacetime everything is finite at an event horizon, and all the way down to either a singularity or a Cauchy horizon.''  The technical content of this theorem is based on showing that the Hadamard form of the  two-point function for a quantum field is not affected by the presence of an event horizon.  (So for a Hadamard quantum state, the renormalized stress-energy tensor [RSET] is automatically finite.)  Unfortunately, for the question we want to raise, the  Fulling--Sweeny--Wald theorem also ``begs the question'', for what it shows is that \emph{if\/} an event horizon forms, \emph{then\/} the quantum field theory is well behaved there.  But this is \emph{not\/} the same as showing that an event horizon will naturally form in semiclassical collapse: A finite but very large contribution from the vacuum polarization term $\langle 0|\hat T_{ab} |0\rangle$ in equation~(\ref{semicl}), while perfectly in agreement with the theorem, would significantly alter the dynamics of collapse. 
In fact, compact horizonless objects, and/or naked singularities, 
are also fully compatible with both the hypotheses and the conclusions of the Fulling--Sweeny--Wald theorem.

\section{Horizons and the choice of the quantum vacuum state}

Our specific proposal for a black hole mimic is ultimately related to the appropriate choice for the quantum vacuum state $|0\rangle$.  Common candidates states are:
\begin{description}
\item[Boulware vacuum:]  
This is singular at any Killing horizon.  The renormalized stress-energy-momentum diverges as $r\to 2m$.

\item[Unruh vacuum:]  
This is designed to be well-behaved at any future Killing horizon, so in particular the renormalized stress-energy-momentum is finite there. 
\end{description}
Neither of these quantum states contain particles in the vicinity of past null infinity $\scri^-$. Accordingly, the key constraint that we shall adopt for our global vacuum state is that it contain no particles in the vicinity of $\scri^-$.  Additionally, far in the past, near past timelike infinity $i^-$, when spacetime 
is static (and nearly Min\-kow\-skian), the quantum state should exhibit properties  qualitatively similar to those of the Boulware vacuum.  This demand is physically appropriate, since the Boulware vacuum is the one that best describes physics in the presence of a static self-gravitating object.  Note that if we were instead to choose a state that in the asymptotic future behaves like the Unruh vacuum, then this would presuppose the formation of a horizon, which however is precisely the issue we wish to investigate.  Such a choice would anyway amount to making  a teleological statement.
Apart from these particular issues, there is an increasing consensus, or at the very least a suspicion, within the general relativity community that event horizons are simply the wrong thing to be looking at.  Sometimes apparent horizons~\cite{textbooks, Lorentzian, apparent} (or better yet, dynamical horizons~\cite{dynamical}, or trapping horizons~\cite{trapping}) are better candidates for characterizing the black hole. (See also~\cite{numerics2, Nielsen, numerics1}.) 

\section{Our specific calculation}

The metric for the spacetime of a spherically symmetric collapsing body can be written in Schwarz\-schild coordinates as
\begin{equation}
\g = - e^{-2\Phi(r,t)} \; (1-2m(r,t)/r)\; \d t^2 + {\d r^2\over 1-2m(r,t)/r} + r^2\left(\d\theta^2+\sin^2\theta\,\d\varphi^2\right)\;.
\end{equation}
In these coordinates an apparent/dynamical/trapping horizon, if it forms, is characterized by
\begin{equation}
{2m(r,t) \over  r} = 1\;.
\end{equation}
In contrast, an event horizon, if it forms, can only be found by back-tracking from future null infinity, $\scri^+$.  

\begin{figure}[!htbp]
\begin{center}
\includegraphics[width=7cm]{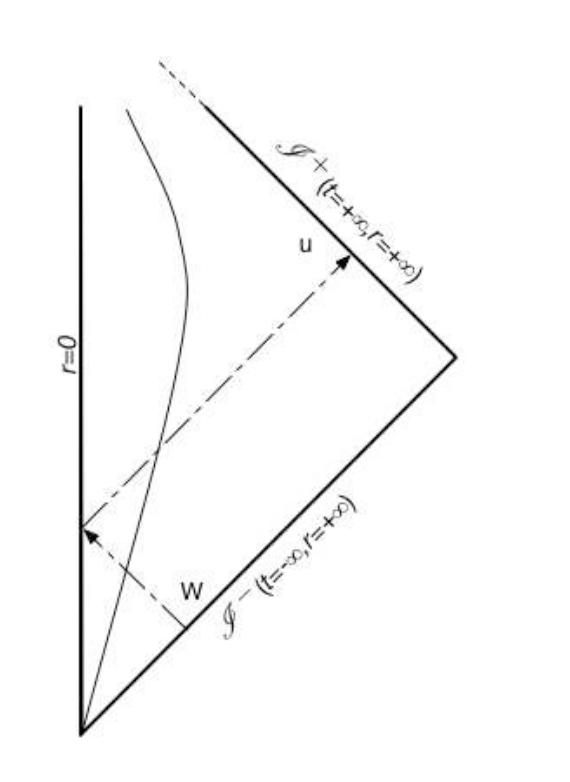}
\caption{Standard Carter--Penrose diagram for the collapse process: Modes ``bounce'' off the centre of the star. We deliberately leave the ``late time'' portion of the diagram vague and ambiguous --- since we do not want to pre-judge what the system settles down to. }
\label{F:bounce}
\end{center}
\end{figure}

In the standard classical conformal Carter--Penrose diagram for the collapse process, figure~\ref{F:bounce}, one truncates the diagram at the centre, $r=0$, and modes of any field residing on the spacetime are said to ``bounce'' off the centre.
A central part of the analysis is then to relate the affine null parameter $W$ on $\scri^-$ to the affine null parameter $u$ on $\scri^+$ via some function $W = p(u)$, which thus encodes a good fraction of all the physics.

\begin{figure}[!htbp]
\begin{center}
\includegraphics[width=9cm]{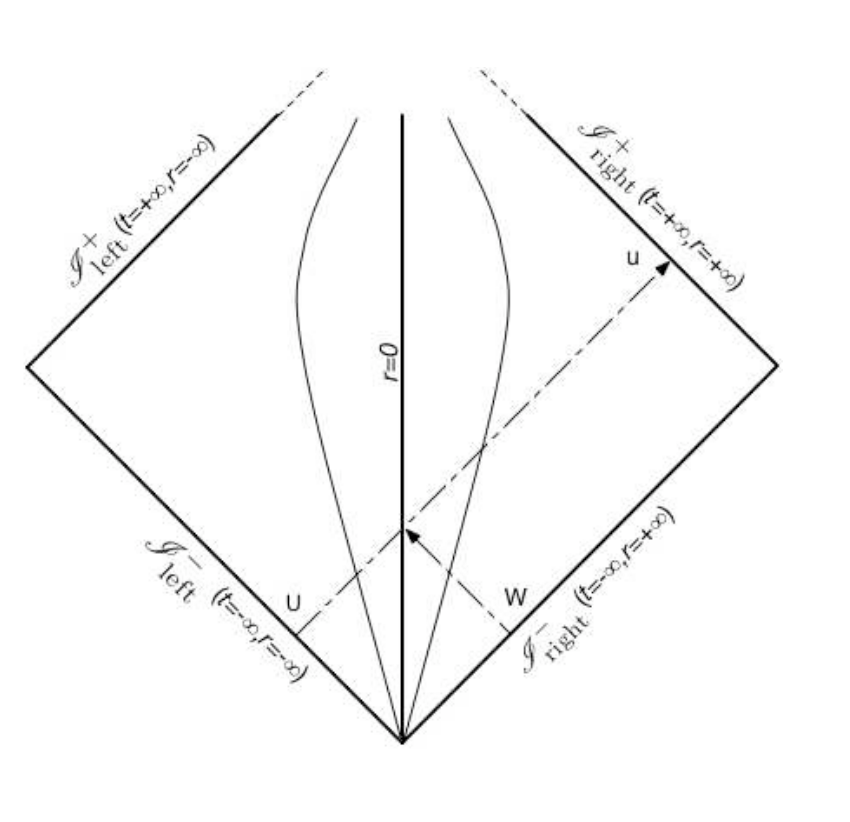}
\caption{Our preferred Carter--Penrose diagram for the collapse process: Modes propagate straight through  the centre of the star. We deliberately leave the ``late time'' portion of the diagram vague and ambiguous --- since we do not want to pre-judge what the system settles down to.  $U$-$W$ coordinates are best in the bottom corner of the diagram, in fact near all of $\scri^-$.  In contrast, $u$-$W$ coordinates are best in the upper-right region, near $\scri^+_\mathrm{right}$.
(And $U$-$w$ coordinates are best in the upper-left region, near $\scri^+_\mathrm{left}$.)}
\label{F:through}
\end{center}
\end{figure}

For technical reasons we prefer to work with a more ``symmetric'' version of the Carter--Penrose causal diagram.  In this version, figure~\ref{F:through}, modes propagate straight through the centre of the collapsing star (located at $r=0$), so that $\scri^-_\mathrm{left}$ is connected to  $\scri^+_\mathrm{right}$, and vice versa.  
Affine null coordinates on $\scri^-$ are now related to affine null coordinates on $\scri^+$ by: 
\begin{equation}
U = p(u)\;; \qquad\qquad  W= p(w)\;.
\end{equation}
Furthermore, we shall work in a $1+1$ dimensional truncation of $3+1$ dimensional spacetime in order to keep the calculation tractable.  It will then be convenient to extend the coordinate $r$ in such a way that on the right-hand part of the diagram $r>0$, while in the left-hand part $r<0$.

For a massless real scalar field, one can expand the field operator using a set of modes that near $\scri^-_\mathrm{left}$ take the form
\begin{equation}
\varphi_\Omega(r,t)\approx\frac{1}{(2\pi)^{3/2}\;(2\Omega)^{1/2}\;|r|}\;\; \e^{-\mathrm{i}\Omega U}\;,
\label{modesonscri-}
\end{equation}
with $\Omega>0$; these are appropriate to define particles in the asymptotic past, before collapse takes place.  Neglecting backscattering, such modes take, near $\scri^+_\mathrm{right}$, the form
\begin{equation}
\varphi_\Omega(r,t)\approx\frac{1}{(2\pi)^{3/2}\;(2\Omega)^{1/2}\;r}\;\;\e^{-\mathrm{i}\Omega p(u)}\;.
\label{modesonscri+}
\end{equation}
However, near $\scri^+_\mathrm{right}$, the modes appropriate for defining particles are not the $\varphi_\Omega$, but others that we denote $\psi_\omega$, with the asymptotic form
\begin{equation}
\psi_\omega(r,t)\approx\frac{1}{(2\pi)^{3/2}\;(2\omega)^{1/2}\;r}\;\;\e^{-\mathrm{i}\omega u}\;.
\label{newmodesonscri+}
\end{equation}
Hence, the state which does not contain particles on $\scri^-$ (defined using the modes $\varphi_\Omega$) turns out to contain particles on $\scri^+$ (defined using the $\psi_\omega$), provided that $p(u)$ is a non-trivial function, such that a $\varphi_\Omega$ mode contains negative-frequency contributions when Fourier-analysed in terms of the $\psi_\omega$.  Defining the $u$-dependent frequency on $\scri^+$,
\begin{equation}
\omega(u,\Omega)=\dot{p}(u)\;\Omega,
\end{equation}
associated with a $\varphi_\Omega$ mode, one can see that mode excitation takes place provided that the adiabatic condition, 
\begin{equation}
{|\dot{\omega}(u,\Omega)|\over \omega^2} \ll 1,
\end{equation}
is violated.  
This occurs for frequencies smaller than 
\begin{equation}
\Omega_0(u)\sim{ |\ddot{p}(u)|\over \dot{p}(u)^2}, 
\end{equation}
which can then be thought of as a frequency marking, at each instant of retarded time $u$, the separation between the modes that have been excited (those with $\Omega\ll\Omega_0$) and those that are still unexcited ($\Omega\gg\Omega_0$).  Intuitively, one may think that there is still an infinite ``reservoir'' of high-energy Boulware-like modes, and that if mode excitation is not sufficiently rapid they will make a potential obstruction to horizon formation.  Indeed, calculations in static models \cite{dfu} show that such modes lead to an energy condition-violating renormalised stress-energy-momentum tensor that diverges as one approaches the horizon.
We feel however, that this is far too na\"ive a picture, being based on results obtained in static spacetimes, and that it can be taken at best as a hint that one should carefully check how the renormalised stress-energy-momentum tensor behaves when the horizon is just about to form.  So, let us now turn to a calculation that takes dynamics explicitly into account.

The spacetime metric can be written using either the set of coordinates $(U,W)$, or $(u,W)$: 
\begin{equation}
\g=-C(U,W)\,\d U\,\d W=-\tilde{C}(u,W)\,\d u\,\d W\;.
\label{metric}
\end{equation}
This gives 
\begin{equation}
C(U,W)={\tilde{C}(u,W)\over \dot{p}(u)},
\label{CC}
\end{equation}
where, for events lying outside the collapsing star, $\tilde{C}(u,W)$ is the metric coefficient of a static spacetime.
%
For any massless quantum field, the renormalised stress-energy-momentum tensor corresponding to a quantum state that behaves like the Boulware vacuum asymptotically in the past has components~\cite{dfu}
\begin{equation}
T_{UU}\propto C^{1/2}\,\partial_U^2\,C^{-1/2}\;,\qquad 
T_{WW}\propto C^{1/2}\,\partial_W^2\,C^{-1/2}\;,\qquad 
T_{UW}\propto C\,R\;,
\label{T}
\end{equation}
where $R$ is the curvature scalar, and the numerical coefficients depend on the specific type of field being considered.  

The component with the most interesting structure is $T_{UU}$.  Using equation (\ref{CC}) and the property $\partial_U=\dot{p}^{-1}\partial_u$ one finds
\begin{equation}
C^{1/2}\,\partial_U^2\,C^{-1/2}=\frac{1}{\dot{p}^2}\left(\tilde{C}^{1/2}\,\partial_u^2\,\tilde{C}^{-1/2}-\dot{p}^{1/2}\,\partial_u^2\,\dot{p}^{-1/2}\right)\;.
\label{key}
\end{equation}
The key point is that the first term within brackets on the right hand side of equation~(\ref{key}) is a static contribution due to the Boulware-like modes, while the second one arises because of the dynamics of collapse.  These two terms, separately, would lead to an arbitrarily large $T_{UU}$ as the horizon is approached, because $\dot{p}$ tends to vanish in this limit.  However, if (and only if) the horizon forms, then the leading contributions of $\tilde{C}^{1/2}\,\partial_u^2\,\tilde{C}^{-1/2}$ and $\dot{p}^{1/2}\,\partial_u^2\,\dot{p}^{-1/2}$ exactly cancel against each other.

For the computation, it is convenient to work in a chart that is regular at the horizon (if it forms), so that the regularity of the stress-energy-momentum tensor can be inferred just by the finiteness of its components.  This is the case for the Painlev\'e-Gullstrand coordinates $(x,t)$, in terms of which the metric is~\cite{pg, analogue, no-trapping}:
\begin{equation}
\g=-c(x,t)^2\,\d t^2+\left(\d x-v(x,t)\,\d t\right)^2\;.
\label{metric-pg}
\end{equation}
A rather technical computation gives~\cite{fate}:
\begin{eqnarray}
T_{tt} &=& U_t^2 \, T_{UU} + 2\, U_t \, W_t \, T_{UW} + W_t^2 \, T_{WW} \\
&=& (c+v)^2\, U_x^2\, T_{UU} - 2\, (c^2-v^2)\, U_x\, W_x\, T_{UW} + (c-v)^2\, W_x^2\, T_{WW} \\
&=& \dot p^2 \, T_{UU} - 2\, \dot p \, T_{UW} + T_{WW}\;;\label{tt}\\
&&\nonumber\\
T_{tx} &=& U_t \, U_x\, T_{UU} + \left(U_t \, W_x + U_x \, W_t\right) \, T_{UW} + W_t \;W_x\; T_{WW} \\
&=& -(c+v)\, U_x^2\, T_{UU} - 2\, v\, U_x\, W_x\, T_{UW} + (c-v)\, W_x^2\, T_{WW} \\
&=&  -{\dot p^2\over c+v} \,T_{UU} + {2\,\dot p\, v\over c^2- v^2} \, T_{UW} + {1\over c-v} \, T_{WW}\;;\label{tx}\\
&&\nonumber\\
T_{xx} &=& U_x^2\, T_{UU} + 2\, U_x\, W_x\, T_{UW} + W_x^2\, T_{WW} \\
&=&
{\dot p^2\,\over (c+v)^2}\,T_{UU} - 2\, {\dot p\over c^2-v^2}\, T_{UW} + {1\over (c-v)^2}\, T_{WW}\;. \label{xx}
\end{eqnarray}
Since at a horizon $c+v\to 0$ and $c-v\to 2c$, the term that presents, potentially, the highest degree of divergence is the one proportional to $T_{UU}$ in $T_{xx}$.  The other potentially dangerous coefficients $\dot{p}^{-2}$ and $\dot{p}^{-1}$ that appear in $T_{UU}$ and $T_{UW}$ are cancelled by corresponding factors in the expressions~(\ref{tt}), (\ref{tx}), and (\ref{xx}).

In the rest of the calculations, we assume that $c(x)=1$ and place the horizon (when it exists) at $x=0$.  Then, assuming that a horizon indeed forms, we can expand $v(x)$ as
\begin{equation}
v(x)=-1+\kappa\,x+{\cal O}(x^2)\;,
\label{vexp}
\end{equation}
where $\kappa$ can be identified with the surface gravity~\cite{analogue, no-trapping}.  The static contribution in equation~\ref{key} is 
\begin{equation}
\tilde{C}^{1/2}\,\partial_u^2\,\tilde{C}^{-1/2}=\frac{\kappa^2}{4}+{\cal O}(x^2)\;,
\label{static}
\end{equation}
and this, taken alone, would cause the $T_{tx}$ and $T_{xx}$ coefficients to diverge.  However, under the hypothesis of horizon formation one also has~\cite{fate}
\begin{equation}
p(u)=U_\mathrm{H}-A_1\,\e^{-\kappa u}+\frac{A_2}{2}\,\e^{-2\kappa u}+\frac{A_3}{3!}\,\e^{-3\kappa u}+{\cal O}(\e^{-4\kappa u})\;,
\label{phorizon}
\end{equation}
where $U_\mathrm{H}$, $A_1>0$, $A_2$, $A_3$ are constants.  The dynamical term in equation~(\ref{key}) is then
\begin{equation}
\dot{p}^{1/2}\,\partial_u^2\,\dot{p}^{-1/2}=\frac{\kappa^2}{4}+\left[-\frac{1}{2}\,\frac{A_3}{A_1}+\frac{3}{4}\left(\frac{A_2}{A_1}
\right)^2\right]\kappa^2\,\e^{-2\kappa u}+{\cal O}(\e^{-3\kappa u})\;.
\label{dynamical}
\end{equation}
Replacing the expressions~(\ref{static}) and~(\ref{dynamical}) into equation~(\ref{key}), one sees that the dominant terms $\kappa^2/4$ cancel against each other, and one remains with a finite contribution that depends on the details of collapse:
\begin{equation}
T_{UU}^\mathrm{assuming~horizon~formation} \propto {1\over \dot p^2}
\left[-\frac{1}{2}\,\frac{A_3}{A_1}+\frac{3}{4}\left(\frac{A_2}{A_1}
\right)^2\right]\kappa^2\;\e^{-2\kappa u}+\dots,
\end{equation}
so that
\begin{equation}
T_{xx}^\mathrm{assuming~horizon~formation} \propto 
\left[-\frac{1}{2}\,\frac{A_3}{A_1}+\frac{3}{4}\left(\frac{A_2}{A_1}
\right)^2\right]\; {1\over x^2}\;\e^{-2\kappa u}+\dots,
\end{equation}
with these expressions holding outside the surface of the collapsing star.
Furthermore, it is easy to realise that this contribution is inversely proportional to the square of the speed at which the collapsing body crosses its gravitational radius.  Hence, for a very slow collapse there is a realistic and concrete possibility that the (energy-condition-violating) renormalised stress-energy-momentum tensor, although finite, could lead to significant deviations from classical collapse when a trapping horizon is just about to form.

In order to reinforce this claim, let us consider a case in which the horizon never forms at any finite time, but is only approached asymptotically in the limit $t\to +\infty$.  In particular, we shall be interested in an exponential approach~\cite{no-trapping}, where the radius of the star depends on time as
\begin{equation}
r(t)=2M+B\,\e^{-\kappa_\mathrm{D} t}\;,
\label{exp}
\end{equation}
with $B$ and $\kappa_\mathrm{D}$ positive constants.  After a brief calculation~\cite{fate, no-trapping} this leads to 
\begin{equation}
p(u)=U_\mathrm{H}-A_1\,\e^{-\kappa_\mathrm{eff} u}\;,
\label{pexp}
\end{equation}
where 
\begin{equation}
\kappa_\mathrm{eff}=\frac{\kappa\kappa_\mathrm{D}}{\kappa+\kappa_\mathrm{D}}<\kappa\;.
\label{keff}
\end{equation}
Here $\kappa_\mathrm{eff}$ can be thought of as a ``reduced surface gravity''.
Interestingly, although no true horizon ever forms, one still gets a Hawking-like flux of Planckian radiation at the temperature~\cite{no-trapping}
\begin{equation}
T = {\kappa_\mathrm{eff}\over 2\pi} = \frac{\kappa\kappa_\mathrm{D}}{2\pi(\kappa+\kappa_\mathrm{D})}.
\end{equation}

Of course, outside the star, the calculation of $\tilde{C}^{1/2}\,\partial_u^2\,\tilde{C}^{-1/2}$ again yields the same result as in equation~(\ref{static}). However, for the second contribution we now have
\begin{equation}
\dot{p}^{1/2}\,\partial_u^2\,\dot{p}^{-1/2}=\frac{\kappa^2_\mathrm{eff}}{4}+{\cal O}(\e^{-2\kappa_\mathrm{eff} u})\;,
\label{dynamical-exp}
\end{equation}
so there is no longer a perfect cancellation between the dominant terms in the static and dynamical contributions.  Indeed, at the leading order, the renormalised stress-energy-momentum tensor outside the star in the limit $x\to 0$ (that is, $r\to 2M$, or $t\to +\infty$) behaves as
\begin{equation}
T_{UU} \propto  \frac{\kappa_\mathrm{eff}^2-\kappa^2}{4\dot p^2} + \dots,
\label{RSET}
\end{equation}
so that
\begin{equation}
T_{xx} \propto \frac{\kappa_\mathrm{eff}^2-\kappa^2}{4\kappa^2\,x^2} + \dots.
\label{RSET2}
\end{equation}
Note that this result is not in contradiction with the Fulling--Sweeny--Wald theorem~\cite{FSW}, because a strict divergence appears only for $t=+\infty$, \ie, at the boundary of spacetime.  However, the renormalised stress-energy-momentum tensor gives an arbitrarily large (albeit finite) energy-condition-violating contribution to the right hand side of the semiclassical Einstein equations as the horizon formation condition $2M/r=1$ is approached.

\section{Implications and Discussion}

\noindent
In the standard collapse, you can argue that the renormalised stress-energy-momentum tensor at horizon-crossing, as felt by infalling matter, is negligible provided:
\begin{enumerate}
\item 
the quantum state is of the Hadamard form (which we have by assertion);
\item
matter is basically freely-falling;
\item
the equivalence principle holds.
\end{enumerate}
The first point tells you that the quantum vacuum has the same ultraviolet form as in Minkowski spacetime, the second point tells you that matter is approximately in a locally inertial frame, and the third point tells you that the local renormalised stress-energy-momentum tensor the matter then ``feels'' must be approximately the same as in Minkowski spacetime --- \ie,  approximately zero (after renormalization).

\begin{figure}[!htbp]
\begin{center}
\includegraphics[width=8cm]{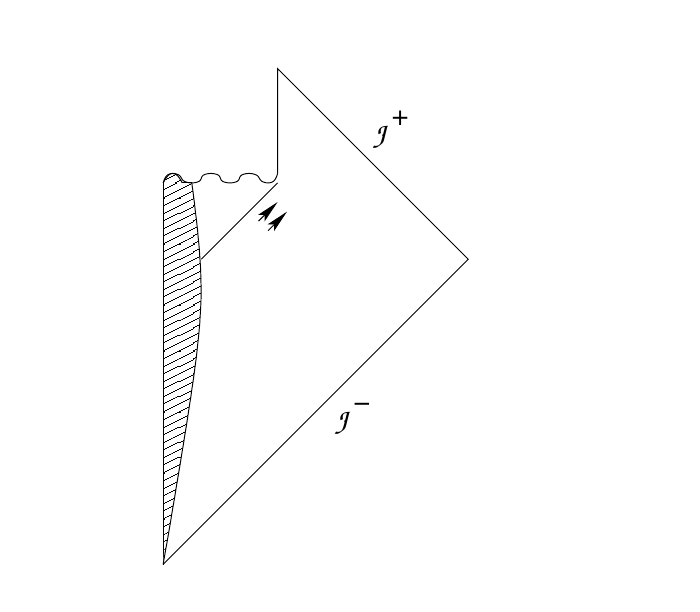}
\caption{Standard Carter--Penrose diagram for an evaporating black hole.  Don't bother asking what happens at the endpoint of the evaporation process --- in the standard causal picture there is no definite answer.}
\label{F:standard}
\end{center}
\end{figure}

In contrast, our result is saying that large deviations from this standard conclusion can arise {\em if\/} matter {\em is not\/} freely falling, but is significantly accelerated (as, by self-consistency, it must be to sustain itself against the gravitational attraction). So we are explicitly violating point~2,  (while we explicitly keep point~1, and implicitly keep point~3).  So if the surface of the star deviates significantly from free-fall, then a large stress-energy-momentum  builds up, which can force it further away from free-fall --- either stopping or exponentially delaying the collapse.  Precisely predicting what happens in a specific collapse scenario relies on extremely messy model-dependent physics.

\begin{figure}[!htbp]%
\vbox{ \hfil \scalebox{0.35}[0.35]{{\includegraphics{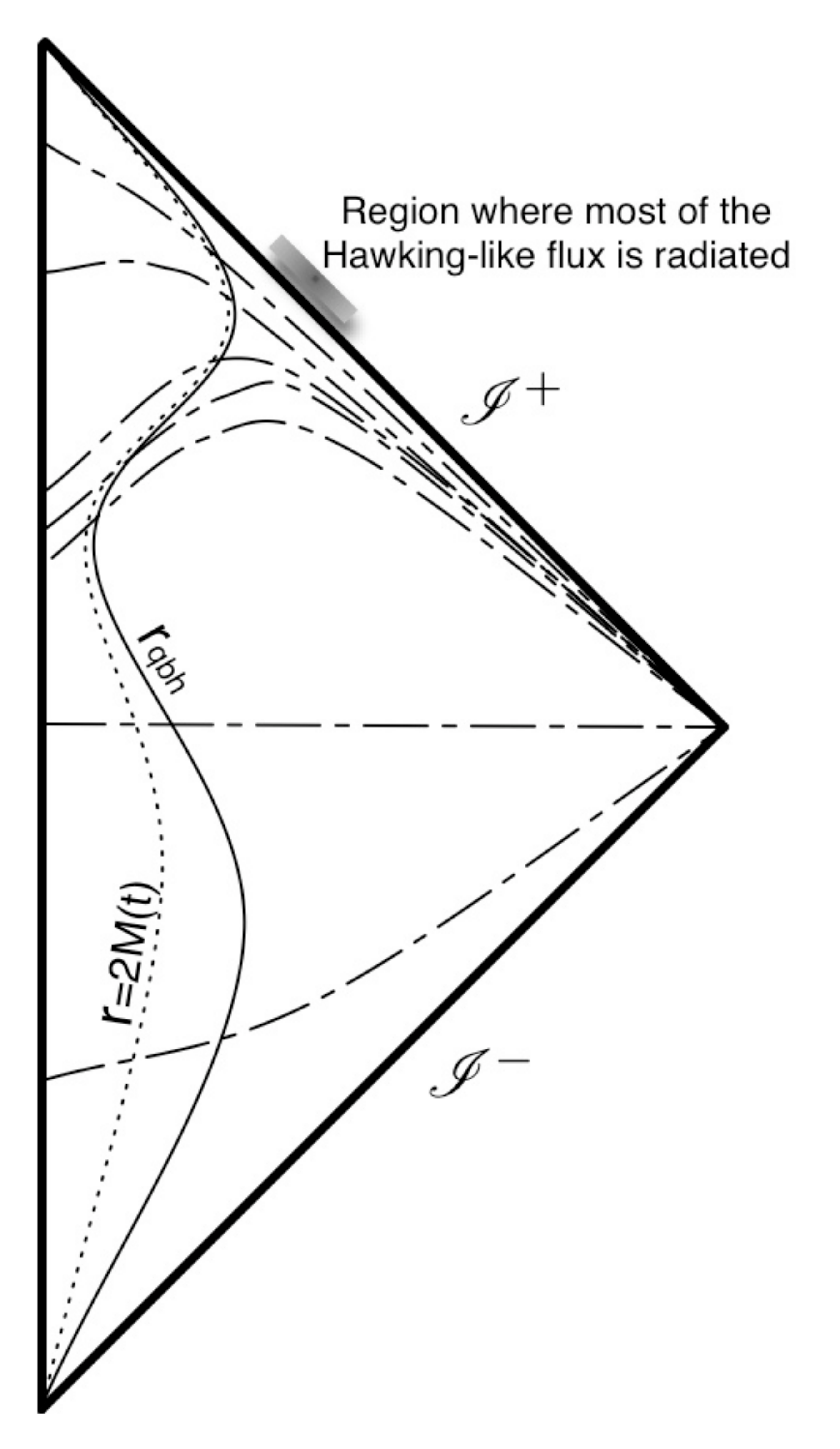}}}\hfil}%
\bigskip%
\caption{Conformal diagram of the spacetime for a quasi-black hole. The solid line represents the surface of the collapsing object; the dotted line is at $r=2M(t)$, where $M(t)$ is the instantaneous mass of the object as measured from $\scri^+$; dashed lines correspond to (Schwarzschild) $t=\mbox{const}$ hypersurfaces. The period of evaporation appears short because of a distortion induced by the representation, but actually corresponds to a very long lapse of time, as one can see from the fact that the lines at $t=\mbox{const}$ crowd around it. This diagram, while nonstandard, is nevertheless compatible with current astrophysical observations of gravitationally active collapse products.}
\label{F:qbh}%
\end{figure}%

Indeed our calculation seems to suggest that if, during the late stages of the collapse, matter is far from free-fall, then a growing RSET can lead to a late time collapse history very different from the classically expected one, possibly leading to a form of asymptotic collapse of the type suggested in~(\ref{exp}). It might even be that this is the solution preferred by nature; this might be due to new particle physics effect coming into place in the late stages of most stellar collapses. In this case the conformal diagram describing the gravitational collapse scenario would not be the standard one of  figure~\ref{F:standard}, but rather that reported in figure~\ref{F:qbh}. This object would then be a ``quasi-black hole'' (not to be confused with the homonymous objects proposed by Lemos and Zaslavskii~\cite{Zaslavskii}, which are static solutions), an object which would not only closely mimic the classical geometry of a black hole, but also, (if the collapse law is exponential at late times), mimic its quantum effects such as Hawking radiation.

\begin{figure}[!htbp]
\begin{center}
\includegraphics[width=9cm]{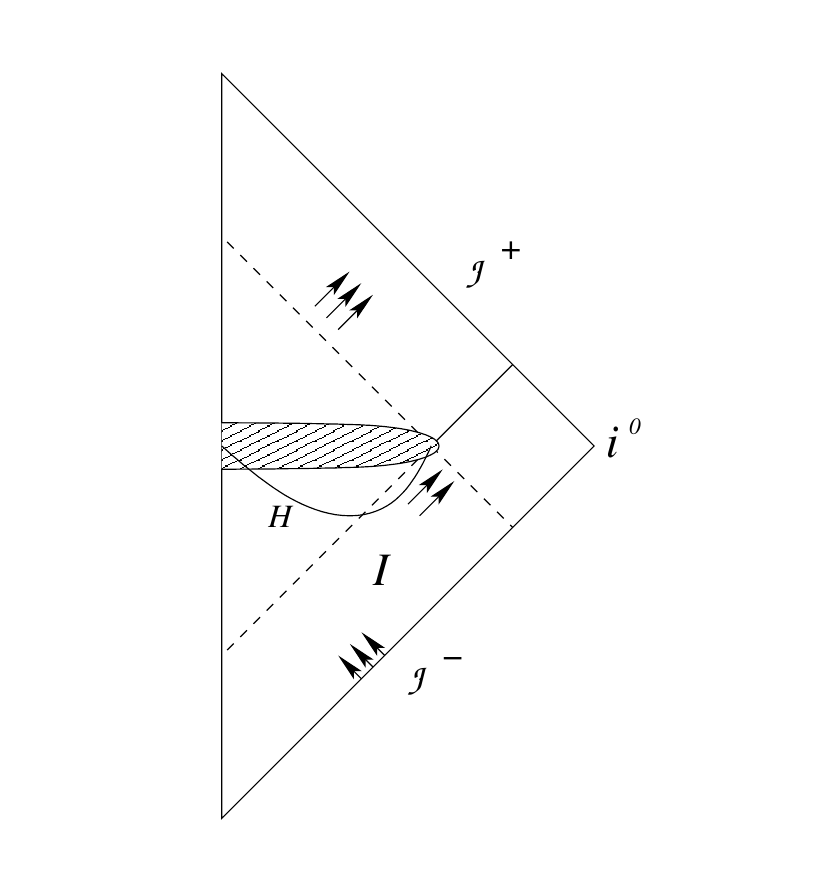}
\caption{Ashtekar--Bojowald version of the Carter--Penrose diagram for an evaporating black hole. The shaded region represents a region of Planck-scale curvature, and possibly large metric fluctuations.}
\label{F:AB}
\end{center}
\end{figure}

\begin{figure}[!htbp]
\begin{center}
\includegraphics[width=7.5cm]{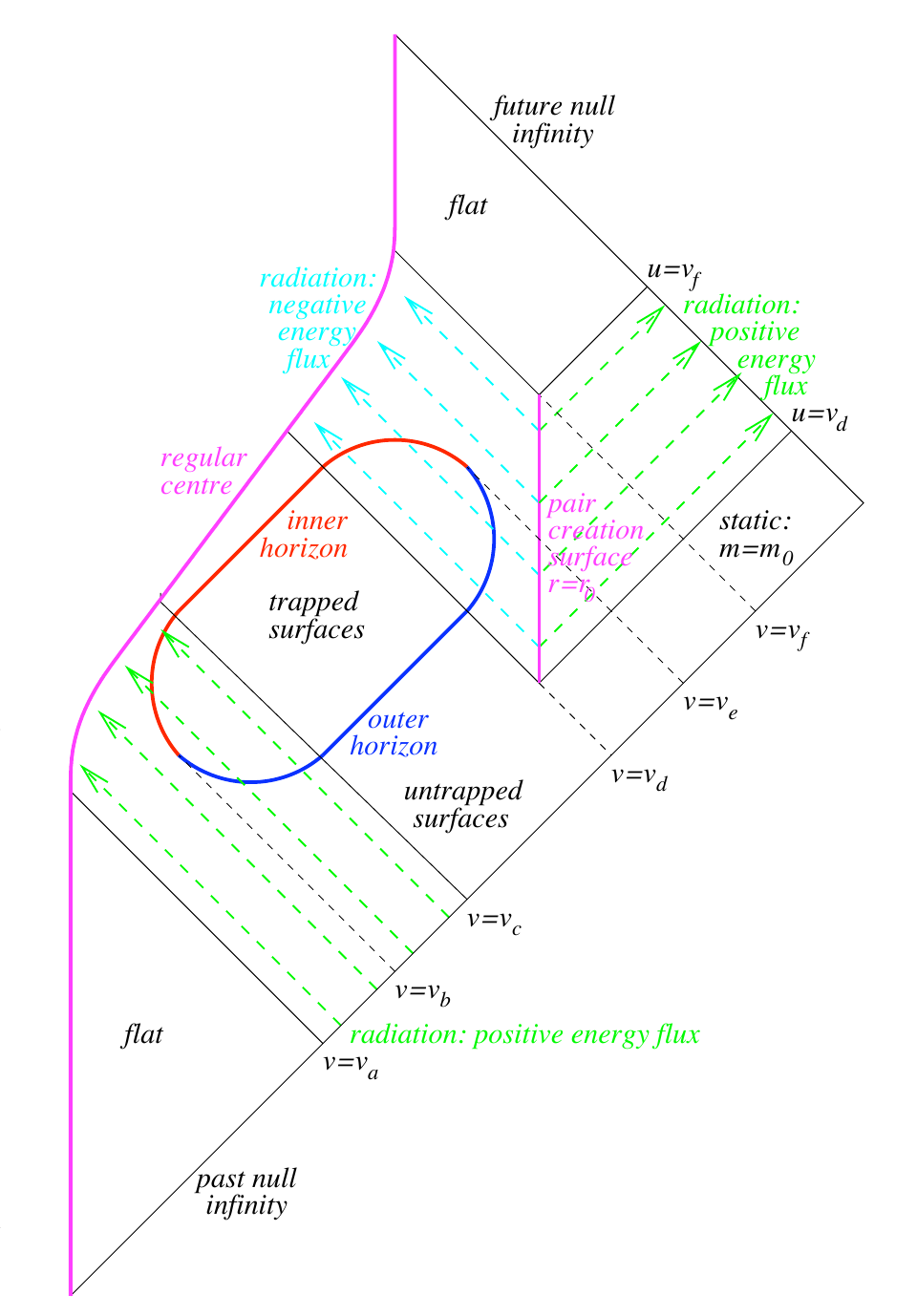}
\caption{Hayward version of the Carter--Penrose diagram for an evaporating black hole.}
\label{F:SH}
\end{center}
\end{figure}

\begin{figure}[!htbp]
\begin{center}
\includegraphics[width=6cm]{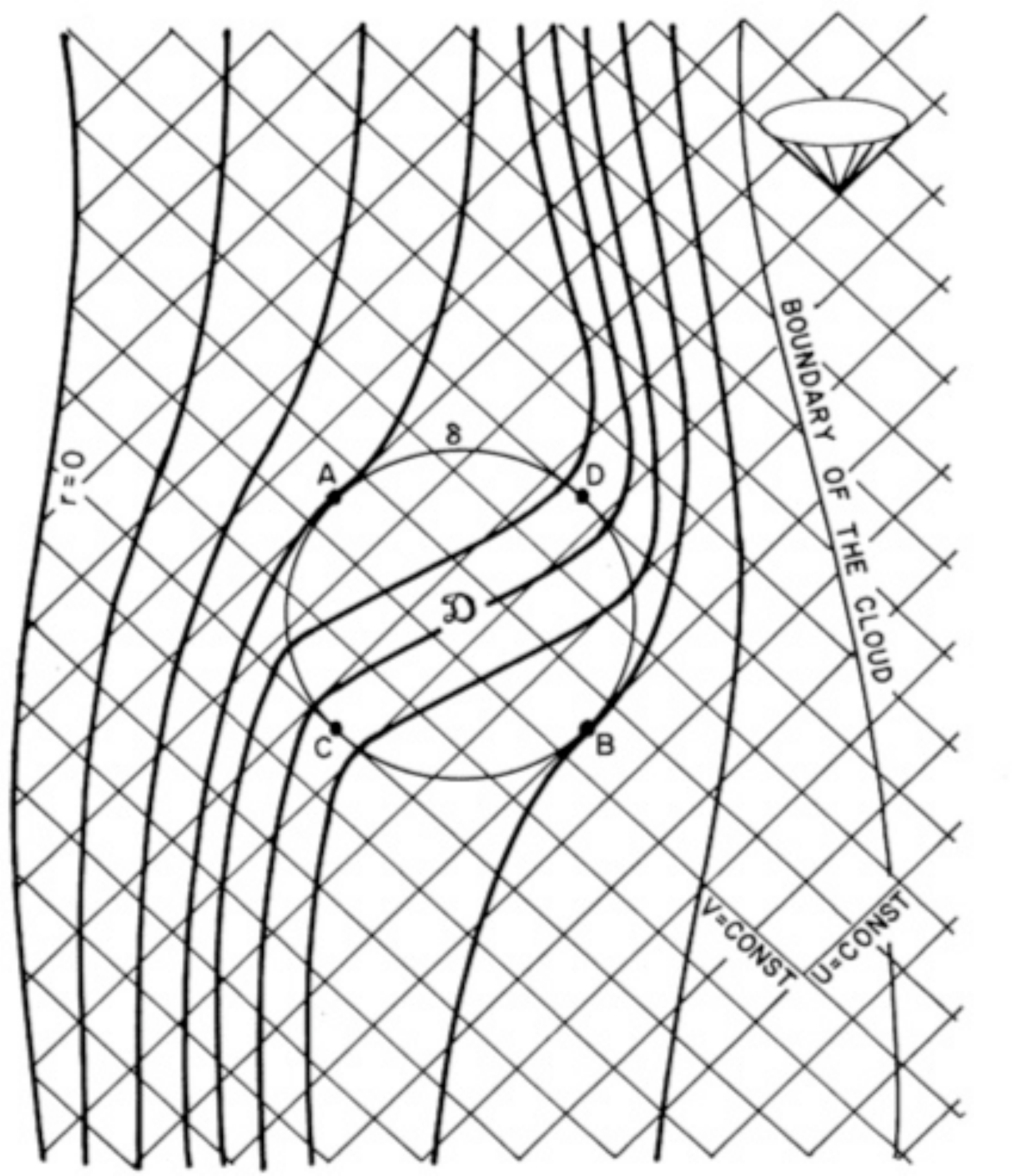}
\caption{Bergmann--Roman double-null causal diagram for a regular collapsing star.}
\label{F:BR1}
\end{center}
\end{figure}

To place our results in a broader perspective: Many physicists are now (for numerous independent reasons) arguing against the standard Carter--Penrose diagram, figure~\ref{F:standard}, for the formation and evaporation of a semi-classical black hole. Apart from the ``physics challenged'',  (whom we shall quietly discount),  there are hints from analogue spacetimes~\cite{no-trapping}, from loop quantum gravity~\cite{Ashtekar-Bojowald}, from string-inspired models~\cite{Amati-et-al},  from attempts at unitarity preservation in our own domain of outer communication~\cite{Stephen}, from one-loop curved-space quantum field theory~\cite{Mottola}, and from abstract studies of the nature of horizons~\cite{Roman}, all hinting at a more subtle history for collapse and evolution. (Canonical versions of alternative causal structures are given by the Carter--Penrose diagrams of figures~\ref{F:AB} and~\ref{F:SH}, and the double-null diagram of figure~\ref{F:BR1}.) Unfortunately,  when attempting to move beyond qualitative statements of this type, specific predictions are frustratingly  model-dependent,  but there is some ``wriggle room'' for interesting new physics.

On a cautionary note, we should point out that several authors have looked at the question of what observational signals for black hole mimics might look like~\cite{Chirenti, Narayan}.  Critically, once you add rotation, the ergosurface is probably more important than the ``would-be horizon''. There is the very real risk of significant ergoregion instabilities~\cite{ergoregion}.

In summary, what our calculation suggests is that it might be possible to have a black hole without having a black hole --- a configuration that is a black hole for (almost) all practical purposes, but might be missing the one key ingredient of having a horizon.  Deep issues of principle remain, and it will be very interesting to see how the whole area of black hole mimics develops over the next few years.


\end{document}